# Ferroelectric Instability under Screened Coulomb Interactions

Yong Wang[*], Xiaohui Liu, J. D. Burton, Sitaram S. Jaswal, and Evgeny Y. Tsymbal[**]

*Department of Physics and Astronomy & Nebraska Center for Materials and Nanoscience,
University of Nebraska, Lincoln, Nebraska 68588, USA*

We explore the effect of charge carrier doping on ferroelectricity using density functional calculations and phenomenological modeling. By considering a prototypical ferroelectric material, $BaTiO_3$, we demonstrate that ferroelectric displacements are sustained up to the critical concentration of 0.11 electron per unit cell volume. This result is consistent with experimental observations and reveals that the ferroelectric phase and conductivity can coexist. Our investigations show that the ferroelectric instability requires only a short-range portion of the Coulomb force with an interaction range of the order of the lattice constant. These results provide a new insight into the origin of ferroelectricity in displacive ferroelectrics and open opportunities for using doped ferroelectrics in novel electronic devices.

Ferroelectric materials are characterized by the spontaneous electric polarization that can be switched between two (or more) orientations.[1] This property makes them attractive for technological applications, such as nonvolatile random access memories, ferroelectric field-effect transistors, and ferroelectric tunnel junctions.[2,3,4] The importance of ferroelectrics also stems from a fundamental interest in the understanding of the electric-dipole ordering, structural phase transitions, and symmetry breaking.[5]

The perovskite $ABO_3$ ferroelectric compounds are especially important group due to the relative simplicity of their atomic structure. The ferroelectric phase transition in these materials is a displacive transition from a high-symmetry paraelectric phase to a polar ferroelectric phase below the critical temperature. This transition is characterized by a decreasing frequency of a transverse optical phonon mode (the soft mode) which drops to zero at the transition point and then becomes imaginary in the ferroelectric phase, corresponding to a collective displacement of ions from their centrosymmetric positions with no restoring force.[6]

The ferroelectric instability can be explained by the interplay between long-range Coulomb interactions favoring the ferroelectric phase and short-range forces supporting the undistorted paraelectric structure.[7] Additional hybridizations between O cation $2p$ and metal anion $d$ orbitals are required to diminish the short-range repulsion and thus to allow for the ferroelectric transition.[8,9] This view is supported by first-principles calculations which indicate that the large destabilizing Coulomb interaction yielding the instability is linked to giant anomalous Born effective charges arising due to the strong sensitivity of O–metal hybridizations to atomic displacements.[10]

While doping a ferroelectric material may enhance its range of functionalities, charge carriers produced by doping will screen the Coulomb interactions that favor the off-center displacements and eventually quench ferroelectricity. This is why it is naturally expected that a ferroelectric phase could not exist in conducting materials. Contrary to this expectation, however, ferroelectric displacements have recently been observed in oxygen reduced conducting $BaTiO_{3-\delta}$.[11,12] It was found that the ferroelectric instability is sustained up to a critical electron concentration $n \approx 1.9\times10^{21}$ cm$^{-3}$, which corresponds to about 0.1 $e$ per unit cell (u.c.) of $BaTiO_3$.

The origin of this "metallic ferroelectricity" is directly related to several important and interesting fundamental questions.[13] How does the screening of the Coulomb interaction affect the ferroelectric displacements? What is the minimum effective range of the Coulomb force to preserve the ferroelectric instability? What happens with the soft mode with charge doping? The answers to these questions would not only provide a better understanding of the nature of ferroelectricity, but also open new possibilities for functional materials.

In this paper, we explore the charge carrier doping effect on ferroelectricity using density functional calculations along with phenomenological modeling based on screened long-range Coulomb interactions and the short-range bonding and repulsion effects. By considering a prototypical ferroelectric material, $BaTiO_3$, we demonstrate that ferroelectric displacements are sustained in electron doped $BaTiO_3$ up to a critical concentration of 0.11 electron per unit cell volume, thus revealing that the ferroelectric phase and conductivity can coexist. Our investigations show that the ferroelectric instability requires only a short-range portion of the Coulomb force with an interaction range of the order of the lattice constant.

Our calculations employ density functional theory (DFT) implemented in the plane-wave pseudopotential code QUANTUM-ESPRESSO.[14] The exchange and correlation effects are treated within the local-density approximation (LDA). The electron wave functions are expanded in a plane-wave basis set limited by a cut-off energy of 600eV. 14×14×14 and 24×24×24 Monkhorst-Pack $k$-points meshes are used for structural relaxation and density of states (DOS) calculations respectively. The self-consistent calculations are converged to $10^{-5}$ eV/u.c. The atomic positions are obtained by fully relaxing the lattice and all the ions in the unit cell until the Hellmann-Feynman force on each atom becomes less than 5 meV/Å. The electron doping in $BaTiO_3$ is achieved by adding extra electrons to the systems with the same amount of uniform positive charges in the background. For the undoped tetragonal $BaTiO_3$, our calculation gives the lattice constant $a = 3.933$Å and $c/a = 1.015$, polarization $P = 28.6$ $\mu C/cm^2$, and Ti-O and Ba-O relative displacements of 0.113Å and 0.091Å respectively, consistent with previous LDA calculations.

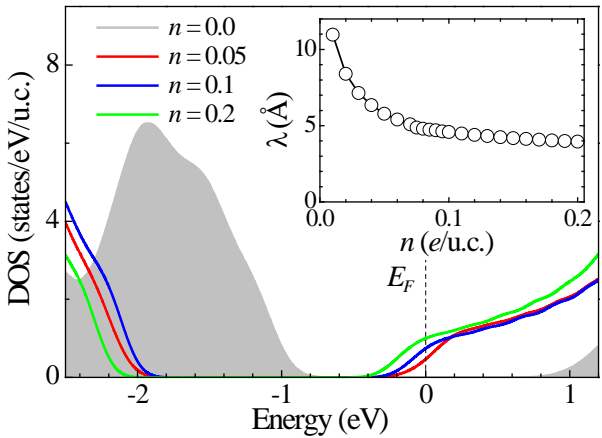

**Fig. 1:** The density of states (DOS) of $BaTiO_3$ for electron doping concentration $n$ = 0.0, 0.05, 0.1 and 0.2 $e/u.c.$. The shaded plot is the DOS of undoped $BaTiO_3$. The vertical dashed line denotes the Fermi energy. The inset shows the Thomas-Fermi screening length $\lambda$ as a function of $n$.

Doping $BaTiO_3$ with electrons pushes the Fermi energy, $E_F$, to the conduction band and screens the electric potential of an ionic charge. Fig. 1 shows the DOS of $BaTiO_3$ for different electron doping concentrations $n$. A typical scale associated with screening is the screening length, $\lambda$, which depends on $n$. We estimate the screening length using the Thomas-Fermi model according to which $\lambda = \sqrt{\varepsilon / e^2 D(E_F)}$. Here $D(E_F)$ is the DOS at $E_F$ and $\varepsilon$ is the dielectric permittivity of undoped $BaTiO_3$ not associated with the spontaneous polarization which we assume to be $\varepsilon \approx 44\varepsilon_0$.[15] Undoped $BaTiO_3$ ($n = 0$) is an insulator so that $D(E_F) = 0$ and hence $\lambda$ is infinite. As $n$ becomes larger, more conduction band states are populated (Fig. 1), thus increasing $D(E_F)$ and reducing the screening length. As seen from the inset in Fig. 1, when $n$ is raised up to 0.2 $e/u.c.$ $\lambda$ decreases down to about 4Å.

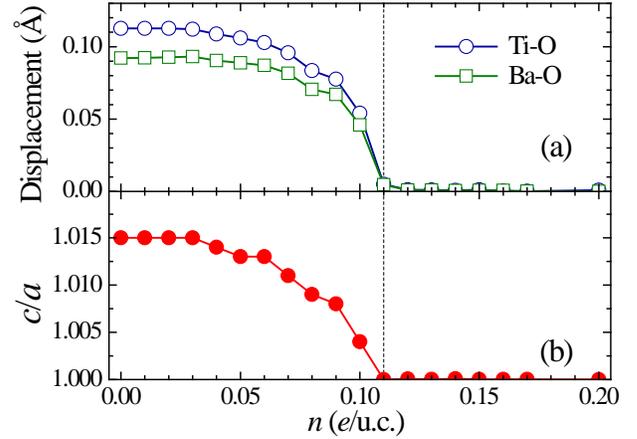

**Fig. 2:** M-O (M = Ti, Ba) relative displacements in $BaTiO_3$ (a) and the ratio of out-of-plane lattice constant $c$ and in-plane lattice constant $a$ (b) as a function of electron doping concentration $n$. The dashed line indicates the critical value $n_c$.

Next we study the effect of screening due to electron doping on the ferroelectric displacements in $BaTiO_3$. Fig. 2a shows the calculated displacements between M and O (M = Ti, Ba) ions as a function of $n$. Surprisingly, we find that ferroelectric displacements hardly change with electron doping up to $n$ as high as $0.05e/u.c.$, and then decay very fast and vanish above the critical electron concentration $n_c = 0.11e/u.c.$. The $c/a$ ratio of $BaTiO_3$ under the increasing $n$, as shown in Fig. 2b, also displays a similar critical behavior as that of polar displacements. $BaTiO_3$ transforms from the tetragonal phase with $c/a = 1.015$ to the cubic phase with $c/a = 1.0$ at $n_c = 0.11e/u.c.$. The critical doping concentration $n_c$ found from first-principles is consistent with the experimental result. According to the inset in Fig. 1 the critical electron concentration $n_c = 0.11e/u.c.$ corresponds to a screening length $\lambda_c \approx 5$Å. Therefore, we conclude that only the short-range Coulomb forces with the interaction range comparable to the lattice constant are responsible for maintaining ferroelectric instability in $BaTiO_3$.

Since changes in hybridization with doping can also affect ferroelectric displacements, we calculate the occupation numbers $N_d$ for the Ti-3$d$ orbitals ($3d_{z^2}$, $3d_{x^2-y^2}$, $3d_{xy}$, $3d_{xz,yz}$) and $N_p$ for the O-2$p$ orbitals of $BaTiO_3$ for different $n$. These occupations reflect the degree of hybridization between Ti-3$d$ and O-2$p$ orbitals. As seen in Fig. 3, $N_d$ decreases and $N_p$ increases very slowly with increasing $n$, so that their change is very small when $n$ is altered from 0 to $n_c = 0.11e/u.c.$. This suggests that the changes in hybridization with doping are negligible. Thus, the dominant mechanism contributing to the ferroelectric critical behavior in $n$-doped $BaTiO_3$ is the screening of Coulomb interactions.

This assertion is further confirmed through our calculations of $p$-doped $BaTiO_3$. Adding holes in $BaTiO_3$

places the Fermi energy in the valence band that is largely determined by the O-2p orbitals. This is different from the n-doped BaTiO$_3$, where the $E_F$ lies in the conduction band built up of the Ti-3d bands. Despite this difference in the bands involved, we find that the p-doped BaTiO$_3$ demonstrates a similar critical behavior of ferroelectric displacements with a critical hole concentration $p_c \approx 0.12$ e/u.c.

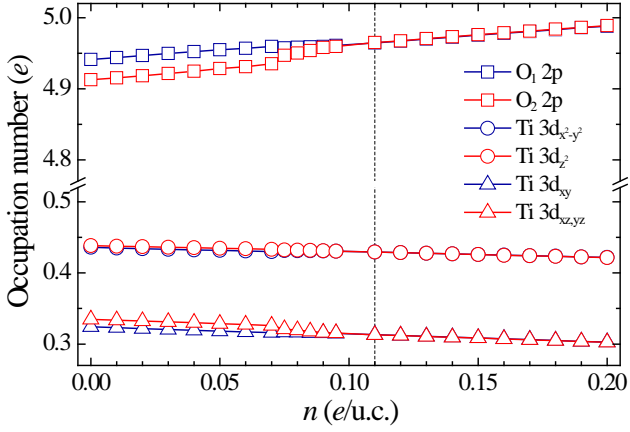

**Fig. 3:** Occupation numbers for Ti-3d and O-2p orbitals as a function of electron concentration n. O$_1$ (O$_2$) correspond to O atoms lying in (off) the TiO$_2$ plane.

The signature of the ferroelectric phase transition can also be seen from the softening of the phonon mode in the paraelectric phase when approaching the critical point with the frequency becoming imaginary in the ferroelectric phase. To confirm the phase transition at the critical concentration we have performed phonon calculations within the density functional perturbation theory, as implemented in QUANTUM-ESPRESSO. In these calculations we consider cubic BaTiO$_3$ with the lattice constant fully relaxed. Fig. 4 shows the lowest frequency of the triple degenerate phonon mode at the Γ point as a function of electron concentration n, along with the relative cation-anion displacements. We see that the frequency remains imaginary up to an electron concentration as high as 0.11e/u.c. and becomes real above this critical concentration. This critical behavior of the ferroelectric instability is echoed by the cation-anion displacements in cubic BaTiO$_3$.

To further understand the critical behavior of ferroelectricity due to screened Coulomb interactions, we have developed a physically realistic model explicitly including the screening effect. We consider a 3-dimensional lattice of ions in the cubic perovskite structure. In the Thomas-Fermi approximation each ion is shrouded by an exponentially decaying screening charge density with screening length λ. The analytical form of the Coulomb interaction energy $w_{ij}$ between two screened point charge $q_i$ and $q_j$ at locations $\mathbf{r}_i$ and $\mathbf{r}_j$, respectively, is $w_{ij}(|\mathbf{r}_i - \mathbf{r}_j|) = q_i q_j w(d)$, where:

$$w(d) = \frac{1}{4\pi\varepsilon_0 d}\left(1 - \frac{d}{2\lambda}\right)e^{-d/\lambda} \quad (1)$$

and $d = |\mathbf{r}_i - \mathbf{r}_j|$ is the distance between the two ions (see Supplementary Material). The factor $(1 - d/2\lambda)e^{-d/\lambda}$ in Eq. (1) is the distance and screening length dependent coefficient, which reflects the effect of screening and converges to 1 as $\lambda \to \infty$. The electrostatic energy per unit cell is given by a lattice sum over all interaction terms of the form (1):

$$W = \frac{1}{2}\sum_{\mathbf{R}}\sum_{i,j=1}^{5} {}' q_i q_j w\left(|\mathbf{r}_i - \mathbf{r}_j + \mathbf{R}|\right), \quad (2)$$

where $\mathbf{R} = a\,(m_x, m_y, m_z)$ are lattice vectors with the m running over all integers. The prime sign on the summation in Eq. (2) indicates that for the $\mathbf{R} = 0$ terms, $i = j$ should be excluded to avoid self-interactions and the factor of ½ takes care of double counting. The summation in Eq. (2) is performed in the spirit of an Ewald sum (see Supplementary Material).

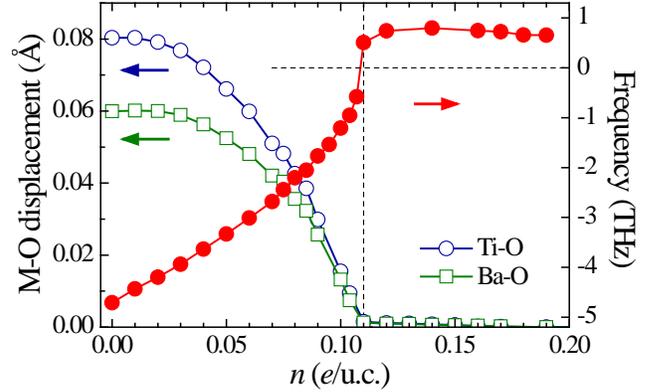

**Fig. 4:** M-O (M = Ti, Ba) relative displacements and phonon frequency of the soft mode at the Γ point in cubic BaTiO$_3$ as a function of electron concentration. Negative sign of frequency indicates an imaginary value of the frequency.

In addition to the long-range electrostatic energy, short-range Ba-O, O-O and Ti-O interactions are also included. These interactions are described by Lennard-Jones potentials $E_0[(R_0/r)^7 - 2(R_0/r)^6]$, along with a O-Ti-O three body potentials given by $k_2(\theta - \theta_0)^2/2$, as parameterized in Ref. 16. The potential parameters are fitted to obtain the same Ba-O and Ti-O displacements in undoped BaTiO$_3$ as those obtained from our DFT calculation. All the parameters of the model except λ are now fixed throughout the calculation.

The total energy of undoped BaTiO$_3$ obtained by adding all the energies described above yields a typical potential with minima at two non-zero polarizations, as seen from the inset in Fig. 5. As the electron screening length $\lambda$ begins to decrease with increasing doping, these minima drop in energy slowly in the beginning. When $\lambda$ approaches the critical value of $\lambda_c$, the two wells become shallower quite rapidly. For $\lambda < \lambda_c$, the wells merge into a single well at $P = 0$ indicating a transition to the paraelectric phase. The critical value predicted by the model, $\lambda_c \approx 5.3$Å, is consistent with that obtained from the Thomas-Fermi estimate based on the DFT calculations. Fig. 5 shows M-O displacements versus the normalized screening length. It is seen that the critical behavior predicted by our model (solid lines) is in agreement with our DFT calculation (open symbols). Thus, our phenomenological model confirms the fact that only a short range portion of the Coulomb interaction is required to sustain ferroelectric displacements.

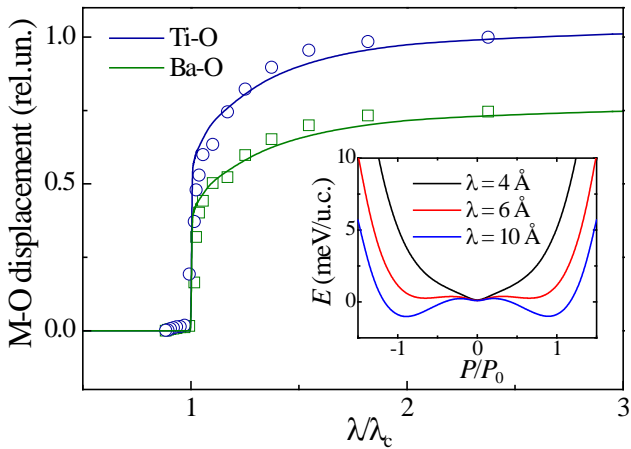

**Fig. 5:** M-O (M = Ti, Ba) relative displacements (in relative units) in cubic BaTiO$_3$ as predicted by the phenomenological model (solid line) and DFT calculation (open symbols). The latter are the same as those in Fig. 4 but plotted versus $\lambda/\lambda_c$ according to the Thomas-Fermi relationship between $\lambda$ and $n$ displayed in the inset of Fig. 1. The inset shows the total energy versus relative polarization for different values of $\lambda$, as follows from the phenomenological model.

The co-existence of the ferroelectric phase and conductivity is interesting for device applications because such a conducting bistable material has new functionalities. Although in such a material an external electric field induces a flow of electric current which makes switching of the ferroelectric displacements difficult, resistive materials may sustain the coercive voltage. For example, ferroelectric tunnel junctions are switchable despite the current flowing across them.[17] Also, there exist means to switch ferroelectrics with no applied voltage which may be used in such devices.[18]

In conclusion, using first-principles calculations and a phenomenological model we have demonstrated that ferroelectric displacements are well preserved in doped BaTiO$_3$ until the doping concentration exceeds a critical value of $n_c = 0.11 e$/u.c. This critical behavior is due to the electron screening of the Coulomb interactions responsible for the ferroelectric instability. The critical screening length is found to be surprisingly small, about 5Å, demonstrating that "short-range" Coulomb interactions are sufficient to lead to collective ferroelectric displacements. This value may be considered as a qualitative estimate for a lower limit for the critical size of BaTiO$_3$ of a few unit cells for the existence of ferroelectricity. Our results provide a new insight into the origin of ferroelectricity in displacive ferroelectrics and open opportunities for using doped ferroelectrics in novel electronic devices.

The authors are thankful to David Vanderbilt and Philippe Ghosez for helpful discussions. This research was supported by the NSF through Nebraska MRSEC (Grant No. DMR-0820521) and Nebraska EPSCoR (Grant No. EPS-1010674). Computations were performed at the University of Nebraska Holland Computing Center.

* Current affiliation: Pacific Northwest National Laboratory, Richland, Washington 99352, USA
** E-mail: tsymbal@unl.edu

# Supplementary Material

Yong Wang, Xiaohui Liu, J. D. Burton, Sitaram S. Jaswal, and Evgeny Y. Tsymbal

*Department of Physics and Astronomy & Nebraska Center for Materials and Nanoscience,*

*University of Nebraska, Lincoln, Nebraska 68588-0299, USA*

**Interaction energy between two screened ions**

In the Thomas-Fermi approximation each ion is shrouded by an exponentially decaying screening charge distribution with screening length $\lambda$. Therefore the potential at **r** generated by a point ion charge at $\mathbf{r}_i$ is

$$\phi_i(\mathbf{r}) = \frac{q}{4\pi\varepsilon_0 |\mathbf{r}-\mathbf{r}_i|} e^{-|\mathbf{r}-\mathbf{r}_i|/\lambda} . \quad (1)$$

We can rewrite this in terms of the Fourier transform of a screened point charge as

$$\phi_i(\mathbf{r}) = \frac{q_i}{(2\pi)^3} \int \tilde{\phi}(k) e^{i\mathbf{k}\cdot(\mathbf{r}-\mathbf{r}_i)} d^3\mathbf{k} . \quad (2)$$

The screened Fourier transform is $\tilde{\phi}(k) = \tilde{\phi}_0(k)/\varepsilon(k)$, where $\tilde{\phi}_0(k)$ is the Fourier transform of the potential of a bare ion with unit charge:

$$\tilde{\phi}_0(k) = \frac{1}{\varepsilon_0 k^2} \quad (3)$$

and $\varepsilon(k)$ is the Thomas-Fermi dielectric function

$$\varepsilon(k) = 1 + \frac{1}{\lambda^2 k^2} . \quad (4)$$

The total screened charge density of this ion is obtained from the Poisson equation as

$$\rho_i(\mathbf{r}) = -\varepsilon_0 \nabla^2 \phi_i(\mathbf{r}) = \varepsilon_0 \frac{q_i}{(2\pi)^3} \int k^2 \tilde{\phi}(k) e^{i\mathbf{k}\cdot(\mathbf{r}-\mathbf{r}_i)} d^3\mathbf{k} . \quad (5)$$

Given a screened point charge $q_j$ at $\mathbf{r}_j$ the work required to bring in another screened point charge $q_i$ from infinity to $\mathbf{r}_i$ is

$$w_{ij}\left(\left|\mathbf{r}_i - \mathbf{r}_j\right|\right) = \int \rho_i(\mathbf{r})\phi_j(\mathbf{r}) d^3\mathbf{r} . \tag{6}$$

Rewriting this integral in terms of the Fourier expressions we obtain

$$w_{ij}\left(\left|\mathbf{r}_i - \mathbf{r}_j\right|\right) = \varepsilon_0 \frac{q_i q_j}{(2\pi)^6} \iiint k^2 \tilde{\phi}(k) e^{i\mathbf{k}\cdot(\mathbf{r}-\mathbf{r}_i)} \tilde{\phi}(k') e^{i\mathbf{k}'\cdot(\mathbf{r}-\mathbf{r}_j)} d^3\mathbf{k}\, d^3\mathbf{k}'\, d^3\mathbf{r} . \tag{7}$$

Therefore, the interaction energy between screened ions $i$ and $j$ separated by distance $d = |\mathbf{r}_i - \mathbf{r}_j|$ can be represented as $w_{ij}\left(\left|\mathbf{r}_i - \mathbf{r}_j\right|\right) = q_i q_j w(d)$, where

$$w(d) = \frac{1}{4\pi\varepsilon_0 d}(1 - \frac{d}{2\lambda})e^{-d/\lambda} \tag{8}$$

which converges to the bare Coulomb potential as $\lambda \to \infty$.

**Evaluation of total electrostatic energy**

The electrostatic energy per unit-cell required to construct the crystal is given by a lattice sum over all interaction terms of the form (8):

$$W = \tfrac{1}{2} \sum_{\mathbf{R}} \sum_{i,j=1}^{5}{}' q_i q_j w\left(\left|\mathbf{r}_i - \mathbf{r}_j + \mathbf{R}\right|\right) \tag{9}$$

Here $\mathbf{R} = a\,(m_x, m_y, m_z)$, are the lattice vectors with the $m$ running over all integers. The ' on the summation over $i, j$ in (9) indicates that for the $\mathbf{R} = 0$ terms, $i = j$ should be excluded to avoid self-interactions and the factor of ½ takes care of double counting.

For large $\lambda$, evaluating (9) via "brute force" summation in real space by truncating those terms with $|\mathbf{R}| > R_{\max}$ is untenable. In the spirit of an Ewald sum, we break up $w(d)$ into two terms: a long range term, $w_L(d)$, which is amenable to summation over a reasonably small

number of Fourier terms, and a short range term, $w_S(d)$, which dies off quickly in real space and therefore is amenable to a reasonably small $R_{max}$, e.g. encompassing only one or two unit-cells.

Explicitly, the Fourier transform of $w(d)$ in (8) is given by

$$\tilde{w}(k) = \varepsilon_0 k^2 \tilde{\phi}(k)^2 = \frac{\lambda^4 k^2}{\varepsilon_0 (\lambda^2 k^2 + 1)^2}. \tag{10}$$

The short range contribution to $w(d)$ comes from Fourier terms with large $k$. Indeed for large $k$, (10) falls off only as $1/k^2$, which gives rise to the singularity in $w(d)$ at $d = 0$. To attenuate these large $k$ contributions out of the Fourier transform and to find the long range contribution $w_L(d)$ to $w(d)$ we multiply (10) by a Gaussian attenuation factor:

$$\tilde{w}_L(k) = \eta \tilde{w}(k) e^{-\sigma^2 k^2} = \frac{\eta \lambda^4 k^2}{\varepsilon_0 (\lambda^2 k^2 + 1)^2} e^{-\sigma^2 k^2}. \tag{11}$$

Here $\eta$ is an as-yet-to-be-determined scaling factor which gives us another degree of freedom to optimally localize the short-range term (more details below) and $\sigma$ is a Gaussian broadening factor roughly corresponding to an effective length of the short-range interaction, which needs to be chosen judiciously to minimize the error between the true expression for $w(d)$ and the approximate $w_S(d) + w_L(d)$. Fourier transforming (11) we find

$$w_L(d) = \frac{1}{(2\pi)^3} \int \tilde{w}_L(k) e^{i\mathbf{k}\cdot\mathbf{d}} d^3\mathbf{k} = \frac{\eta \lambda^4}{2\pi^2 \varepsilon_0 d} \int_0^\infty \frac{e^{-\sigma^2 k^2} \sin(kd)}{(\lambda^2 k^2 + 1)^2} k^3 dk. \tag{12}$$

The short-range contribution $w_S(d)$ is obtained straightforwardly:

$$w_S(d) = w(d) - w_L(d). \tag{13}$$

Using (12), the leading order terms of $w_s$ as $d$ tends toward infinity we obtain

$$w_S(d \to \infty) \approx \frac{(\eta e^{\sigma^2/\lambda^2} - 1)}{8\pi\lambda\varepsilon_0} e^{-d/\lambda} - \frac{(\sigma^2 \eta e^{\sigma^2/\lambda^2} + \lambda^2 \eta e^{\sigma^2/\lambda^2} - \lambda^2)}{4\lambda^2 \pi \varepsilon_0 d} e^{-d/\lambda} + \frac{\eta\sigma}{2\pi^{3/2}\varepsilon_0 d^2} e^{-d^2/4\sigma^2}. \tag{14}$$

By choosing $\eta$ so that the first term in (14) is zero, we obtain

$$\eta = e^{-\sigma^2/\lambda^2}. \tag{15}$$

Using (15), the full expressions for $w_S(d)$ and $\tilde{w}_L(k)$ are given by

$$w_S(d) = \frac{(2\lambda^2 + 2\sigma^2 - d\lambda)}{16\pi\varepsilon_0 \lambda^2 d} \operatorname{erfc}\left(\frac{d}{2\sigma} - \frac{\sigma}{\lambda}\right) e^{-d/\lambda}$$
$$+ \frac{(2\lambda^2 + 2\sigma^2 + d\lambda)}{16\pi\varepsilon_0 \lambda^2 d} \operatorname{erfc}\left(\frac{d}{2\sigma} + \frac{\sigma}{\lambda}\right) e^{d/\lambda} - \frac{\sigma^2}{4\pi\varepsilon_0 \lambda^2 d} e^{-d/\lambda}, \tag{16}$$

$$\tilde{w}_L(k) = \frac{e^{-\sigma^2/\lambda^2} \lambda^4 k^2}{\varepsilon_0 (\lambda^2 k^2 + 1)^2} e^{-\sigma^2 k^2}.$$

Now we return to (9) and approximate it in terms of the long and short range Ewald contributions:

$$W' = W_L - W_{self} + W_S = \tfrac{1}{2}\sum_{\mathbf{R}}\sum_{i,j=1}^{5} q_i q_j w_L\left(|\mathbf{r}_i - \mathbf{r}_j + \mathbf{R}|\right) - \tfrac{1}{2}\sum_{i=1}^{5} q_i^2 w_L(0)$$
$$+ \tfrac{1}{2}\sum_{\mathbf{R}}\sum_{i,j=1}^{5}{}' q_i q_j w_S\left(|\mathbf{r}_i - \mathbf{r}_j + \mathbf{R}|\right). \tag{17}$$

Since we have removed the singularity at $d = 0$ from $w_L(d)$ we can rewrite $W_L$ without the $'$ by subtracting away the terms for $i = j$ when $\mathbf{R} = 0$ which sum to give rise to the self-interaction term

$$W_{self} = \tfrac{1}{2}\sum_{i=1}^{5} q_i^2 w_L(0), \tag{18}$$

where

$$w_L(0) = -\frac{(3\lambda^2 + 2\sigma^2)}{8\pi\varepsilon_0 \lambda^3}\operatorname{erfc}\left(\frac{\sigma}{\lambda}\right) + \frac{(\lambda^2 + \sigma^2)}{4\pi^{3/2}\varepsilon_0 \sigma} e^{-\sigma^2/\lambda^2}. \tag{19}$$

Writing $w_L$ in its Fourier transform, $W_L$ expressed in reciprocal space can be derived as

$$W_L = \tfrac{1}{2} \sum_{\mathbf{G}} \sum_{i,j=1}^{5} q_i q_j \frac{1}{(2\pi)^3} \int \tilde{w}_L(k) e^{i\mathbf{k}\cdot(\mathbf{r}_i-\mathbf{r}_j)} \delta(\mathbf{k}-\mathbf{G}) d^3\mathbf{k}$$
$$= \frac{1}{2a^3} \sum_{\mathbf{G}} \tilde{w}_L(G) \sum_{i,j=1}^{5} q_i q_j e^{i\mathbf{G}\cdot(\mathbf{r}_i-\mathbf{r}_j)} = \frac{1}{2a^3} \sum_{\mathbf{G}} \tilde{w}_L(G) S(\mathbf{G}),$$
(20)

where we have defined the structure factor

$$S(\mathbf{G}) = \sum_{i,j=1}^{5} q_i q_j e^{i\mathbf{G}\cdot(\mathbf{r}_i-\mathbf{r}_j)}.$$
(21)

and $\mathbf{G}$ are the reciprocal lattice vectors: $\mathbf{G} = (2\pi/a)(n_x, n_y, n_z)$, where the $n$ runs over all integers up to a maximum cut-off of $N_{max}$.

By matching the approximate electrostatic energy $W'$ to the true electrostatic energy $W$, which can be calculated via brute force for a few representative structures and screening lengths, we find a maximum error less than 0.1meV for $N_{max} = 7$ and $\sigma = 0.6$ Å.